\RequirePackage{lineno} 
\documentclass[letterpaper,twocolumn,showpacs,amsmath,amssymb,prc,nofootinbib]{revtex4}

\usepackage{graphicx}

\usepackage{bm}

\usepackage{epsfig}
\usepackage{amssymb}
\usepackage{graphicx,color}
\usepackage{dcolumn}
\usepackage{tabularx}

\begin{document}

\title{Experimental upper bound and theoretical expectations for parity-violating neutron spin rotation in $^{4}$He}
\author{H. E.~Swanson}
\affiliation{University of Washington and Center for Experimental Nuclear Physics and Astrophysics, Box 354290, Seattle, WA 98195, USA}
\author{C. D.~Bass}
\affiliation{LeMoyne College, 1419 Salt Springs Road, Syracuse, NY 13214, USA}
\author{T. D.~Bass}
\affiliation{Indiana University and Center for the Exploration of Energy and Matter, 2401 Milo B. Sampson Lane, Bloomington, IN 47408, USA}
\author{B. E.~Crawford}
\affiliation{Gettysburg College, 300 North Washington Street, Gettysburg, PA 17325, USA}
\author{J. M.~Dawkins}
\affiliation{Indiana University and Center for the Exploration of Energy and Matter, 2401 Milo B. Sampson Lane, Bloomington, IN 47408, USA}
\author{K.~Gan}
\affiliation{The George Washington University, 2121 I Street N.W., Washington, D.C. 20052, USA}
\author{B. R.~Heckel}
\affiliation{University of Washington and Center for Experimental Nuclear Physics and Astrophysics, Box 354290, Seattle, WA 98195, USA}
\author{J. C.~Horton}
\affiliation{Indiana University and Center for the Exploration of Energy and Matter, 2401 Milo B. Sampson Lane, Bloomington, IN 47408, USA}
\author{C.~Huffer}
\affiliation{North Carolina State University, 2401 Stinson Drive, Raleigh, NC 27695, USA }
\author{D.~Luo}
\affiliation{Indiana University and Center for the Exploration of Energy and Matter, 2401 Milo B. Sampson Lane, Bloomington, IN 47408, USA}
\author{D. M.~Markoff}
\affiliation{North Carolina Central University/TUNL, 1801 Fayetteville Street, Durham, NC 27707, USA}
\author{A. M.~Micherdzinska}
\affiliation{The George Washington University, 2121 I Street N.W., Washington, D.C. 20052, USA}
\author{H. P.~Mumm}
\affiliation{National Institute of Standards and Technology, 100 Bureau Drive, Gaithersburg, MD 20899, USA}
\author{J. S.~Nico}
\affiliation{National Institute of Standards and Technology, 100 Bureau Drive, Gaithersburg, MD 20899, USA}
\author{M.~Sarsour}
\affiliation{Georgia State University, 29 Peachtree Center Avenue, Atlanta, GA 30303-4106, USA}
\author{E. I.~Sharapov}
\affiliation{Joint Institute for Nuclear Research, Joliot-Curie 6, 141980 Dubna, Russia}
\author{W. M.~Snow}
\affiliation{Indiana University and Center for the Exploration of Energy and Matter, 2401 Milo B. Sampson Lane, Bloomington, IN 47408, USA}
\author{S. B.~Walbridge}
\affiliation{Indiana University and Center for the Exploration of Energy and Matter, 2401 Milo B. Sampson Lane, Bloomington, IN 47408, USA}
\author{V.~Zhumabekova}
\affiliation{Al-Farabi Kazakh National University, Al-Farabi Ave. 71, 050038 Almaty, Kazakhstan}


\date{\today}
\begin{abstract}

Neutron spin rotation is expected from quark-quark weak interactions in the Standard Model, which induce weak interactions among nucleons that violate parity. We present the results from an experiment searching for the effect of parity violation via the spin rotation of polarized neutrons in a liquid $^{4}$He medium.  The value for the neutron spin rotation angle per unit  length in $^{4}$He,  $d\phi/dz =(+2.1 \pm 8.3 (stat.)\, ^ {+2.9} _{-0.2} (sys.))\times10^{-7}$\,rad/m, is consistent with zero. The result agrees with the best current theoretical estimates of the size of nucleon-nucleon weak amplitudes from other experiments and with the expectations from recent theoretical approaches to weak nucleon-nucleon interactions. In this paper we review the theoretical status of parity violation in the $\vec{n}+^{4}$He system and discuss details of the data analysis leading to the quoted result.  Analysis tools are presented that quantify systematic uncertainties in this measurement and that are expected to be essential for future measurements.

\end{abstract}


\maketitle

\section{Introduction}\label{sec:intro}

Neutron spin rotation is expected from quark-quark weak interactions in the Standard Model, which induce weak interactions among nucleons that violate parity.  Because the energies involved in our measurement are well below the energy scale $\Lambda_{QCD}$ where quantum chromodynamics (QCD) becomes a strongly-interacting theory, the nucleon-nucleon (NN) weak interaction involves the unsolved nonperturbative limit of QCD. The expected size of the parity-odd rotation angle in low energy NN interactions is about $10^{-6}$\,rad/m to $10^{-7}$\,rad/m~\cite{Stodolsky1974}. The measurement presented in this work has achieved a precision in this regime. 

The hadronic weak interaction (HWI) in general and the NN weak interaction amplitudes in particular are scientifically interesting for several reasons~\cite{Musolf2006,Erler2005,Snow2005,Nico2005,Liu2007}. Because the range for $W$ and $Z$ exchange between quarks is small compared to the nucleon size, HWI are first-order sensitive to quark-quark correlations in hadrons. This is also true for strangeness-changing nonleptonic weak decays of hadrons. Both nonleptonic weak kaon decays (which have been known for decades to be greatly amplified in the $\Delta I=1/2$ channel) and nonleptonic weak decays of hyperons exhibit deviations from the relative sizes of weak amplitudes expected by chiral symmetry whose dynamical source is still not fully understood~\cite{Donoghue86}. If these unexpected patterns in the isospin dependence of nonleptonic weak amplitudes are confirmed by measurements in the NN and few nucleon systems, it would indicate that this dynamical puzzle operates for all light quarks (rather than just the strange quark) and is therefore a nontrivial QCD ground state dynamical phenomenon of general interest~\cite{Musolf2006}.  The weak NN interaction is also one of the few systems thought to be sensitive to quark-quark neutral current effects at low energy because charged currents are generically suppressed in $\Delta I = 1$ NN processes  by ${V_{us}^{2}/V_{ud}^{2}} \simeq 0.1$~\cite{Adelberger1985,Haxton2013}. Quark-quark and NN weak interactions also induce parity-odd effects in electron scattering~\cite{Bec01, Beck01, Beck03, Beise2005}, nuclear decays~\cite{Adelberger1985}, compound nuclear resonances~\cite{Bowman1993,Tomsovic2000}, and atomic structure, where they are the microscopic source for nuclear anapole moments~\cite{Zel57,Fla80,Flambaum1984,Woo97,Haxton2001,Haxton2002,Tsi09,Sahoo2016,Dzuba2017}.

QCD possesses only vector interactions and its gauge symmetry is unbroken in its low temperature phase, and in this phase QCD is therefore expected to conserve parity (although it is suspected that QCD can spontaneously break parity symmetry in high temperature phases~\cite{Kh98}).  The residual parity-violating HWI is therefore expected to be induced only by
quark-quark weak interactions as described in the Standard Model. There are two model-independent statements that one can make about this interaction: one at the quark level for energies above $\Lambda$ and the other at the nucleon level for energies below $\Lambda$. For energies above $\Lambda$ but below the electroweak scale, the quark-quark weak interaction can be written in a current-current form with pieces that transform under (strong) isospin as $\Delta I=0,1,2$. At the nucleon level for energies below $\Lambda$, one can show that five independent weak transition amplitudes are present in NN elastic scattering at low energy~\cite{Danilov1965}: the $\Delta I = 1$ transition amplitudes between $^3S_1 -{^3P_1}$ and $^1S_0 -{^3P_0}$ partial waves; the $\Delta I = 0$ transition amplitudes between $^3S_1 -{^1P_1}$ and $^1S_0 -{^3P_0}$ partial waves; and the $\Delta I = 2$ transition amplitude between the $^1S_0 -{^3P_0}$ partial waves. Unfortunately, it is not yet possible to perform a quantitative calculation in the Standard Model to interpolate between these two regimes.

The relative strengths of the different four-quark operators just below the electroweak scale evolve under the QCD renormalization group and can be calculated in QCD perturbation theory~\cite{DAI,KAPLAN} between the electroweak scale and $\Lambda$. The unsolved nonperturbative QCD dynamics have so far prevented theorists from extending these calculations below $\Lambda$ to make direct contact with low energy NN weak amplitudes. If one wants to probe the nonperturbative physics of the ground state of an asymptotically-free gauge theory like QCD, an interaction that is weak, perturbative, calculable at short distance, and does not itself significantly affect the strong dynamics is the type of probe one wants to employ. The development of quark-quark weak interactions into NN weak interactions as the distance scale increases satisfies these criteria. It is in this sense that measurements of the NN weak interaction can be thought of as an \lq\lq inside-out\rq\rq probe of the ground state of QCD.

Theoretical work on the HWI can be organized into three broad classes depending on how the strong interaction dynamics are treated: (1) model-dependent approaches that posit a specific dynamical mechanism for the interaction, (2) model-independent approaches with a direct connection to QCD based on its symmetries, and (3) direct calculation from the Standard Model. Model-dependent approaches include meson exchange, QCD sum rules~\cite{HENLEY, Lobov2002, HWANG}, nonlocal chiral quark models~\cite{LEE}, SU(3) Skyrme models~\cite{Meissner1999}, and models motivated by the recent nonperturbative treatment of QCD based on the AdS/CFT correspondence~\cite{GAZIT}. In the meson exchange picture the NN weak interaction is modeled as a process in which the three lightest mesons $(\pi,~\rho,~{\rm and}~\omega)$ couple to one nucleon via the weak
interaction at one vertex and to the second nucleon via the strong interaction at the other vertex. An attempt to calculate the weak meson-nucleon couplings of the HWI from the Standard Model using a valence quark model for QCD was first made by Desplanques, Donoghue, and Holstein (DDH) in 1980~\cite{Des80} and updated in 1998~\cite{Des98}. In the DDH model HWI observables are expressed in terms of six weak meson-nucleon coupling constants: $h^1_\pi$,$h^0_\rho$,$h^1_\rho$,$h^2_\rho$,$h^0_\omega$,$h^1_{\omega}$, where the subscript indicates the exchange meson and the superscript labels the isospin change. The results obtained by DDH have served as a de facto benchmark for experimental and theoretical work in the field for several years.  An experimental program was outlined and the calculations specifying the relation between the corresponding observables and the weak coupling constants were reviewed, compiled, and in some cases performed by Adelberger and Haxton in 1985~\cite{Adelberger1985}.

More recently a model-independent theoretical framework has spurred renewed theoretical interest and experimental effort. This framework is based on effective field theory (EFT) methods that have been applied with success to low energy processes in the meson and nucleon sectors and have now been extended to describe the HWI. It has the advantage of
being, by construction, the most general theoretical description consistent with the symmetries and degrees of freedom of low energy QCD, and it involves within this framework a perturbative expansion in the small parameter $p/\Lambda$, where $p$ is a typical internal momentum involved in the reaction. Because most planned experiments to resolve NN
weak interaction effects occur in this energy range, one can imagine determining the unknown couplings of the operators in the EFT description from experiment. The theory takes different forms depending on the treatment of strong interaction effects and whether or not pions are treated as separate dynamical degrees of freedom. For processes in which the momentum transfers involved are below $\approx 40$\,MeV a pionless EFT~\cite{Phil09, Schindler2013} which treats both the strong and weak interactions consistently in an EFT framework and possesses five parameters in the low energy limit labeled by the partial wave transition amplitudes is appropriate. For higher momenta it becomes important to include explicit pion degrees of freedom~\cite{Musolf2006,Zhu2005,Liu2007}. The weak NN interaction has been analyzed using chiral effective  field theory~\cite{DeVries2013, DeVries2016}, which can also treat both the strong and electroweak interactions on the same footing. This approach was used in conjunction with the existing pp parity violation data to suggest a value for $h_{\pi}^{1}$ in reasonable agreement with the result later reported by NPDGamma~\cite{Blyth2018}. The chiral EFT approach also has the potential to treat parity violation in nuclear few body systems like $n+^{3}$He and $n+^{4}$He.  

Of the five independent weak transition amplitudes present in NN elastic scattering at low energy~\cite{Danilov1965, Danilov1972}, two are now fixed from experiment:  the $^{1}S_{0} -{^3}P_{0}$ proton-proton amplitude~\cite{Potter74, Balzer80, Kistryn87, Eversheim91} and the $^{3}S_{1}\rightarrow^{3}P_{1}$ amplitude from the parity-violating 2.2\,MeV gamma-ray asymmetry $A_{\gamma}$ in polarized slow neutron capture on protons~\cite{Blyth2018}. It is not possible in the foreseeable future to determine the weak NN interaction solely using measurements in the NN system. Fortunately, strong interaction effects are now calculable~\cite {Pieper01} in few-body nuclei and weak amplitudes can be added as a perturbation, so it is possible to constrain NN weak amplitudes using measurements of parity violation in few-body nuclei. The experimental strategy to determine the weak NN interaction therefore employs parity-odd observables in light nuclei such as H, D, $^{3}$He, and $^{4}$He. If one wants to use pionless EFT in these systems, one must also ensure that the internal momenta are small enough that the expansion parameter of EFT is still small. The effects of possible three-body parity-odd interactions have been estimated and found to be negligible~\cite{Schindler2010}.

The possibility to calculate the weak NN amplitudes on the lattice was analyzed long ago~\cite{Bea02} and is now under active investigation. The most easily accessible amplitude for lattice calculations is the $\Delta I = 2$, $^1S_0 -{^3P_0}$ channel because this amplitude does not possess disconnected quark loop diagrams, which are computationally expensive on the lattice. An effort to calculate parity violation in this partial wave on the lattice was listed as a \lq \lq grand challenge\rq\rq\  problem in exoscale computing~\cite{Ex2009}, and the first pioneering lattice gauge theory calculation of the $\Delta I = 1$, $^3S_1 -{^3P_1}$ channel using large pion masses and other approximations was performed~\cite{Wasem2012}. Recent work relating this amplitude via a chiral theorem to a four-quark operator contribution to the neutron-proton mass difference~\cite{Guo2018,Guo2019} may make it possible to perform a reliable lattice calculation despite the presence of the disconnected quark diagrams in this case. If both of these lattice efforts succeed, then two of the five low-energy pionless EFT parameters will be determined from the Standard Model. In combination with the parallel efforts to calculate the $\Delta I = 1/2$ and $\Delta I = 3/2$ amplitudes on the lattice in nonleptonic kaon decay~\cite{Christ09}, the success of these efforts would offer the exciting possibility of a direct comparison of nontrivial nonleptonic weak interaction amplitudes with the Standard Model.

Recently the $1/N_{c}$ expansion of QCD~\cite{tHooft1974, Witten1979} which correctly reproduces the relative strengths of strong NN couplings and other low energy QCD observables at typically the 20-30 percent level~\cite{Jenkins1998, Cohen2012, DeGrand2016} has been applied to the NN weak interaction sector. Phillips {\it et al.}~\cite{Phillips2015, Samart2016} constructed the $1/N_{c}$ expansion of the seven couplings in the meson exchange model, and Schindler {\it et al.}~\cite{Schindler2016} produced this expansion in the EFT approach. Gardner {\it et al.}~\cite{Gardner2017} use this work in combination with NN weak data in two and few nucleon systems to argue that all of the existing data seems to be consistent with the $N_{c}$ dominance of two of the 5 NN weak amplitudes: the $\Delta I = 2$, $^1S_0 -{^3P_0}$ amplitude and a linear combination $\Lambda^{+}_{0}$ of the $\Delta I = 0$ amplitudes proportional to $3/4 (^3S_1 -{^1P_1})+1/4 (^1S_0 -{^3P_0})$. More experimental data is needed to confirm these $1/N_{c}$ arguments, but for now they are useful as the best guidance for where one might look for large parity-violating (PV) effects. We will present the implications of this estimate for parity-odd neutron spin rotation in $n+^{4}$He in Section~\ref{sec:theory}.

The remainder of this paper is organized as follows. In Section~\ref{sec:theory} we discuss the phenomenon of parity-odd neutron spin rotation and the existing theoretical work on neutron spin rotation in $^{4}$He. Section~\ref{sec:exp} summarizes the experimental method, results of the polarimeter's characterization, and experimental uncertainty, all of which are covered in more detail in Ref.~\cite{Snow2015}.  In Section~\ref{sec:systematics} we describe how measurements are used to estimate overall systematic uncertainty.  In Section~\ref{sec:anal}, we present the data analysis, and Section~\ref{sec:uncert} discusses the experimental uncertainties. Lastly, the results are given in Section~\ref{sec:concl}. These results were first reported in a short paper in this journal~\cite{Snow2011} and were later used to constrain possible exotic parity-odd interactions of the neutron from light axial vector boson exchange~\cite{Yan2013} and from gravitational torsion~\cite{Lehnert2014} and nonmetricity~\cite{Lehnert2017}.  A recent paper~\cite{Snow2015} discusses the experimental method, apparatus, and the sources of some possible systematic effects.
 
\section{$\vec{n}+^{4}\rm He$ Parity-odd Neutron Spin Rotation Theory}
\label{sec:theory}

As described in detail in~\cite{Snow2015}, the phenomenon of parity-violating neutron spin rotation can be described in terms of neutron optics~\cite{Michel1964}. A rotation of the tip of the neutron polarization about its momentum vector as it passes through isotropic, unpolarized matter describes a corkscrew in space, thereby manifestly violating mirror symmetry. From a neutron optical viewpoint, this phenomenon is caused by the presence of a helicity-dependent neutron index of refraction. The index of refraction $n$ of a medium in terms of the coherent forward scattering amplitude $\lim _{{\vec q} \to 0} f({\vec q})=f(0)$ for a low-energy neutron is

\begin{equation}
n=1-\frac{2\pi\rho f(0)}{ k^{2}},
\end{equation}

\noindent where $\vec {k}$ is the incident neutron wave vector and $\rho$ is the number density of scatterers in the medium. At low energy in an unpolarized medium, $f(0)$ is the sum of two terms: a parity-conserving term $f_{PC}$ dominated by the strong interaction and consisting of s-waves at low energy, and a parity-violating term $f_{PV}$ that contains only weak interactions and is dominated by a p-wave contribution at low energy

    \begin{equation}
    f(0)=f_{PC}+f_{PV}(\vec{\sigma}_{n}\cdot\vec{k}_{n}),
    \label{PCPVScatterAmp}
    \end{equation}

\noindent where $\vec{\sigma_{n}}$ is the neutron spin vector, which has opposite signs for the positive and negative longitudinally-polarized neutron spin-states. As a neutron moves a distance $z$ in the medium, the two longitudinal polarization states accumulate different phases: $\phi_{\pm}=\phi_{PC} \pm \phi_{PV}$. $\phi_{PV}$ causes a relative phase shift of the two neutron longitudinal polarization components and therefore a rotation of the neutron polarization about its momentum

    \begin{equation}
    \phi_{PV}=\varphi_{+}-\varphi_{-}=2\varphi_{PV}=4\pi\rho{z}f_{PV}.
    \label{PVrotAng}
    \end{equation}

\noindent The scattering amplitude $f_{PV}$ can be written in terms of  the neutron mass, $M$, and the weak matrix element formed from the incoming and outgoing wave functions, $\psi_i$ and $\psi_o$, and the weak Hamiltonian

   \begin{equation} 
    f_{PV}=-\frac{M}{2\pi}Re<^{4}He,\psi_{i}\mid{H_{wk}}\mid\psi_{o},^{4}He>.
    \label{fPV}
    \end{equation}

Because the parity-odd amplitude is proportional to $k$, the rotary power per unit length $d\phi/dz=4\pi \rho f_{PV}$ tends to a constant for low energy neutrons~\cite{Stodolsky1974}. An order-of-magnitude estimate leads one to expect weak rotary powers in the range of $10^{-6}$\,rad/m to $10^{-7}$\,rad/m. In the case of parity violation in compound resonances in neutron-nucleus reactions there are amplification mechanisms~\cite{Forte78} that can amplify parity-odd observables by factors as large as $10^{5}$. Parity-odd neutron spin-rotation at cold neutron energies has been observed in $^{117}$Sn~\cite{Forte80}, Pb~\cite{Heckel82}, and $^{139}$La~\cite{Heckel84}, and has been searched for in $^{133}$Cs, Rb, and $^{81}$Br~\cite{Saha1990}. All of the nonzero PV spin rotation observations so far seen with meV energy neutron beams come from larger effects in compound nuclear resonances from the tails of a higher-energy p-wave resonance. In the case of Pb, there is still some controversy about which isotope is enhanced~\cite{Bolotsky1996,Lobov2000,Lobov2002,Andrzejewski2004}.

The parity odd amplitudes involved in $\vec{n}+^{4}$He spin rotation can be treated to an excellent approximation as perturbations in this and all low energy NN weak interaction processes. They can be estimated in meson-exchange ({\it i.e.}, DDH) or other QCD models, parametrized using EFT, or calculated from the Standard Model using lattice gauge theory. In the meson-exchange model, a HWI observable $(O_{PV})$ can be expressed completely in terms of six weak meson-nucleon coupling constants $O_{PV} = a^1_\pi h^1_\pi + a^0_\rho h^0_\rho + a^1_\rho h^1_\rho + a^2_\rho h^2_\rho + a^0_\omega h^0_\omega + a^1_\omega h^1_{\omega} $. The coefficients $a^{\Delta  I}$ are determined from theoretical calculations, where $\Delta I $ is the change in isospin. The only existing calculation in the literature (to our knowledge) of $d\phi/dz$  in $\vec{n}+^{4}$He ($d\phi/dz = -0.97 f_\pi -0.22 h^0_\omega +0.22 h^1_\omega -0.32h^0_\rho +0.11h^1_\rho$) was conducted within the DDH framework~\cite{Dmitriev1983}. In that approach, $d\phi/dz$ in $\vec{n}+^{4}$He spans a range of $\pm 1.5 \times 10^{-6}$\,rad/m using the original DDH ranges for the couplings. This broad range of possibilities is dominated by the uncertainties in the weak couplings, which reflect in part our poor understanding of quark-quark correlation physics in QCD and whose determination is the major goal of the experimental work. There is also an additional layer of theoretical uncertainty involved in the correct calculation of the linear combination of the NN weak couplings to which a given $P$-odd NN observable is sensitive.

The best estimate for $d\phi / dz$  now comes from the recent analysis of the implications of the $1/N_{c}$  expansion as applied to the weak NN interaction~\cite{Phillips2015,Schindler2016}. Within this framework, parity violation in $\vec{n} +^{4}$He spin rotation is predicted to be $d\phi/dz = (9 \pm 3)  \times 10^{-7}$\,rad/m. This estimate uses the following elements: (1) the new result for $f_\pi = (2.6 \pm 1.2) \times 10^{-7}$ from the measurement of the parity-odd gamma asymmetry in polarized slow neutron capture on protons from the NPDGamma collaboration~\cite{Blyth2018}; (2) the value $h^0_\rho =(-36 \pm 8) \times 10^{-7}$ from previous measurements of parity violation in proton-proton scattering; and (3) the  $1/N_{c}$ scaling relations among the couplings, which imply that the combined effect of all of the rest of the terms in the expression for $d\phi/dz$ gives only a few percent correction to the dominant  $h^0_\rho$ contribution.  The final result from a measurement of parity violation in the $\vec{n}+^{3}$He reaction conducted at the Spallation Neutron Source at Oak Ridge National Laboratory is being prepared for publication~\cite{Crawford2019}, and this result could help to check this estimate. The estimate assumes that one can take the results of the 1983 calculation of Dmitriev {\it et al.}~\cite{Dmitriev1983} at face value and that the large internal momenta inside the $^{4}$He nucleus do not significantly affect the analysis. More theoretical work should establish whether these assumptions are accurate enough to preserve this prediction.

A calculation using Greens Function Monte Carlo techniques~\cite{Nollett2011} could greatly improve the precision of the relative weighting with which the different amplitudes contribute. A calculation of $\vec{n} +^{4}$He spin rotation performed in the effective field theory framework has not yet appeared. A calculation of parity violation in neutron spin rotation in $^{4}$He using the Fadeev-Yakubovsky nonrelativistic few body formalism including two and three body interactions was recently completed~\cite{Lazauskas2019}. In comparison to the Dmitriev {\it et al.} calculation, it has a smaller weak pion contribution and results in an expected spin rotation angle slightly larger than the $(9 \pm 3) \times 10^{-7}$\,rad/m quoted above. Two new measurements of the low energy s-wave scattering length for neutrons on $^{4}$He, which is important input for any calculation of neutron spin rotation as a check on the strong interaction component of the calculation, have been conducted and are being analyzed~\cite{Haddock2018,Huber2018}.

\section{Experimental Method}
\label{sec:exp}

\begin{figure}
\begin{center}
\includegraphics[width=8cm]{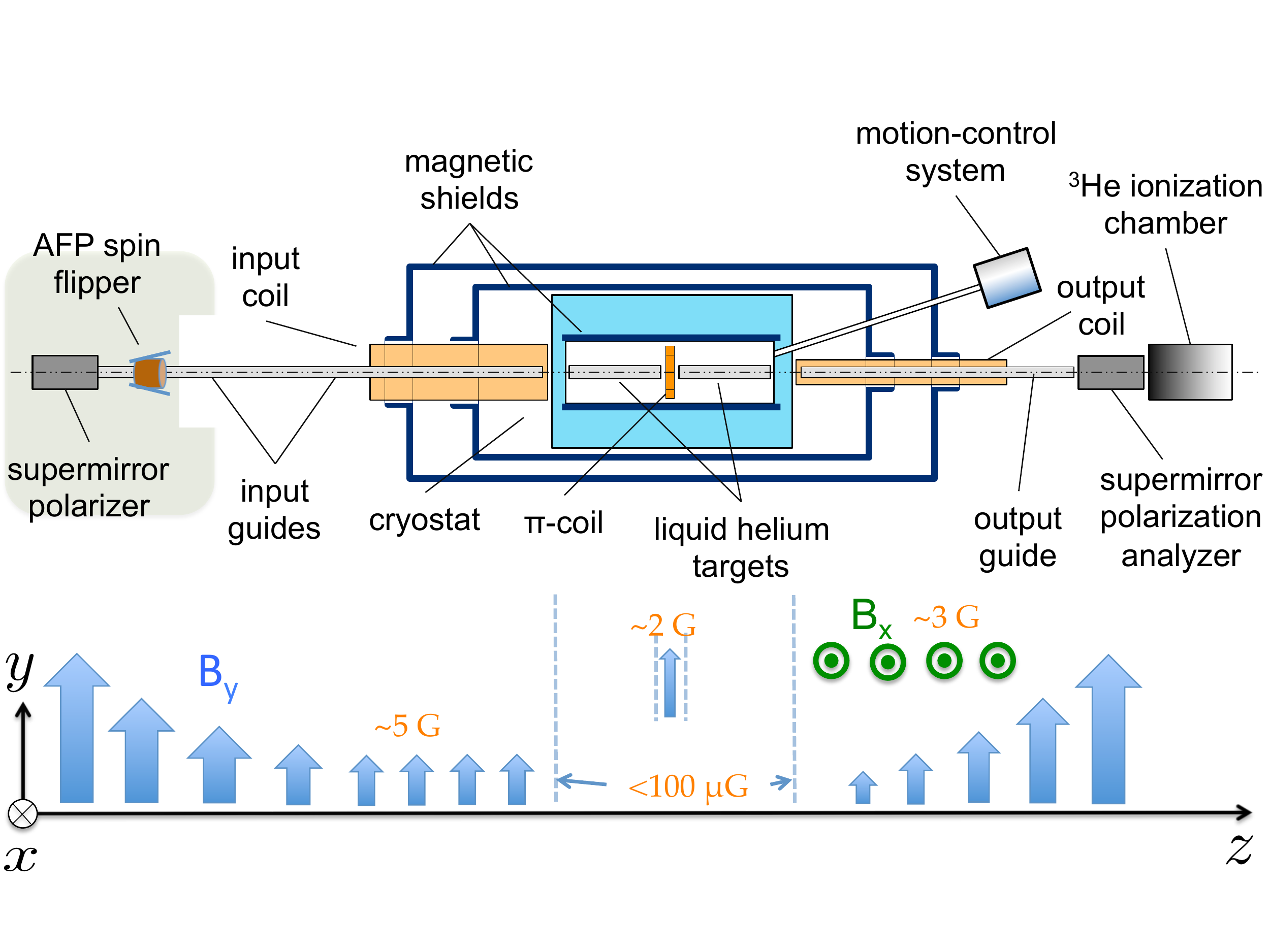}
\caption{Top view of an apparatus to measure PV neutron spin rotation in liquid helium~\cite{Snow2015}.  Typical magnetic field directions and magnitudes are given at the bottom of the figure.}
\label{fig:apparatus}
\end{center}
\end{figure}

The experimental technique has been presented in other papers~\cite{Bass09, Snow2009, Snow2011, Snow2015} and only an overview is given here. The measurement was performed at the NIST Center for Neutron Research, which operates a 20\,MW, D${}_{2}$O-moderated research reactor that provided thermal neutrons to nine experimental stations and cold neutrons by moderation in liquid hydrogen~\cite{Rush2011}. Neutrons from the NG-6 neutron guide at the cold neutron source~\cite{NicoJRES05} are vertically polarized by a supermirror neutron polarizer and then transported by vertical magnetic fields followed by a nonadiabatic transition to the low-field target region (see Fig.~\ref{fig:apparatus}).  The neutron spin rotation apparatus functions as a neutron analog of a crossed polarizer-analyzer optics experiment.  Neutrons reaching the end of the target region pass through another non-adiabatic magnetic transition to a region of horizontal field which preserves any horizontal components of rotation acquired in the target region. The field in the output coil slowly twists about the longitudinal axis to adiabatically rotate the horizontal component of any acquired rotation of the polarization $\pm 90^{\circ}$ into the vertical direction either parallel or anti-parallel to the vertically aligned supermirror analyzer.  A $^3$He ionization chamber then detects neutrons transmitted through the analyzer.  

To measure the neutron spin rotation, the experiment counts neutrons from a crossed polarizer ($\hat{x}$) and analyzer ($\hat{y}$) system, in which the final analyzer direction alternates between $+\hat{y}$ and $-\hat{y}$. By flipping the direction of the vertical field in the output coil, one forms an asymmetry in the neutron counts to determine the rotation angle in the target region. The rotation angle is then given by
 
\begin{equation}
    \sin{\phi}=\frac{1}{PA}\frac{N_{+}-N_{-}}{N_{+}+N_{-}} ,
    \label{Asymmetry}
\end{equation}

\noindent where $N_+$ and $N_-$ are the transmitted neutron counts for the two directions of the analyzer, and $PA$ is the analyzing power of the  crossed polarizer-analyzer pair. The directions and relative sizes of the transport fields are indicated in Fig.~\ref{fig:apparatus}.

\begin{figure}
\begin{center}
\includegraphics[width=8cm]{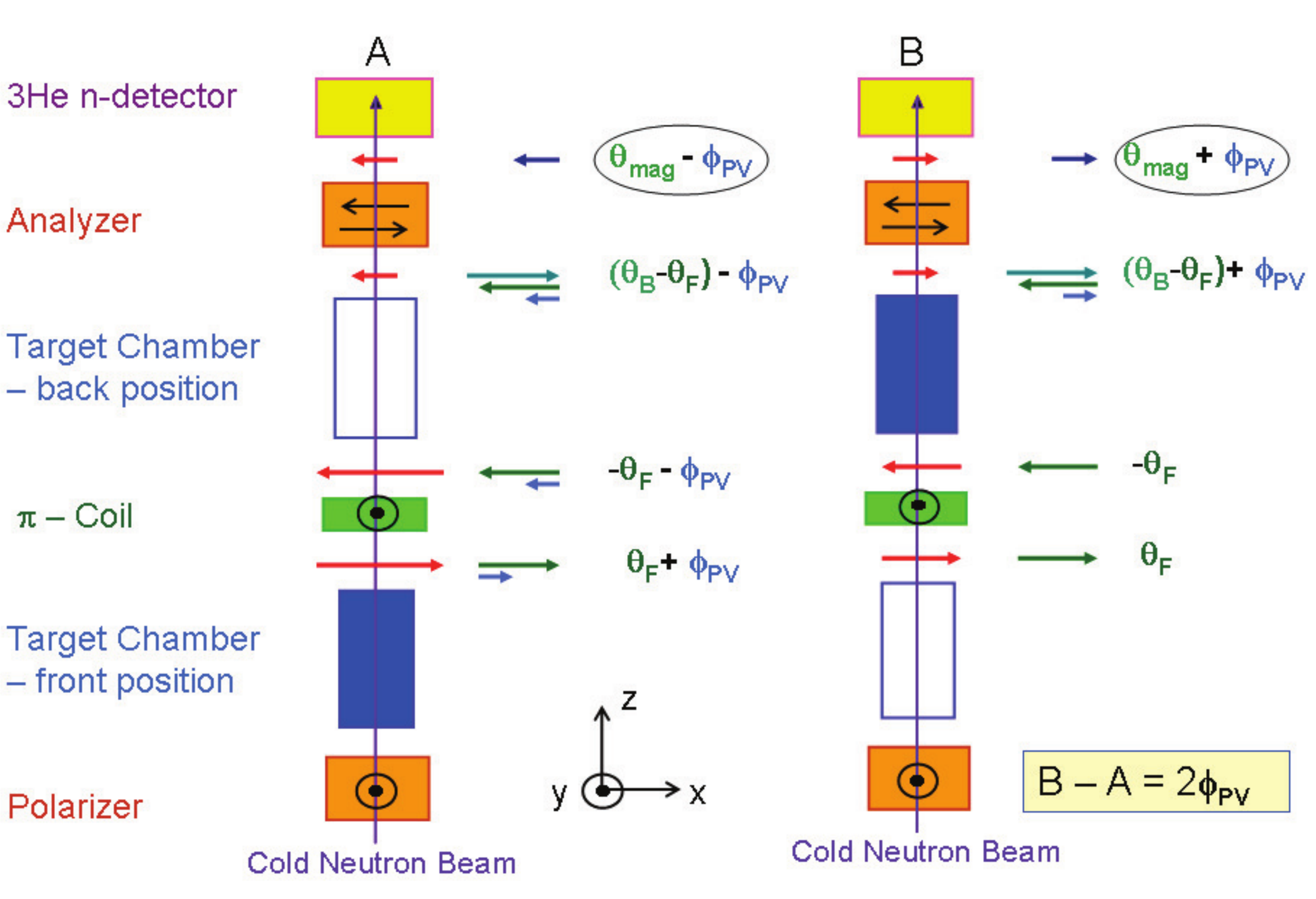}
\caption{Conceptual diagram (top view) illustrating the strategy to isolate the PV neutron spin rotation signal in the presence of a large background from residual longitudinal-magnetic-fields~\cite{Snow2015}. The arrows show the projection of the neutron polarization-vector onto the plane of the figure at different points along the apparatus and for two different target states A and B, which correspond to the liquid present in the upstream and downstream chamber, respectively.}
\label{target_idea}
\end{center}
\end{figure}

The central challenge for a neutron polarimeter measuring rotations at the $10^{-7}$\,rad/m level is to distinguish the parity-violating rotations of interest from the much larger rotations due to ambient magnetic fields.  Thus the apparatus is designed to suppress magnetic fields, cancel rotations from magnetic fields {\it in situ}, and further cancel these rotations through subtraction in a series of measurements involving target motion. Magnetic fields in the target region are kept in the 10\,nT range by two layers of room-temperature magnetic shielding and an additional layer of Cryoperm~\footnote{Certain commercial equipment, instruments, or materials are identified in this paper in order to specify the experimental procedure adequately. Such identification is not intended to imply recommendation or endorsement by the National Institute of Standards and Technology, nor is it intended to imply that the materials or equipment identified are necessarily the best available for the purpose.} at 4\,K in the target cryostat.  To cancel rotations from residual magnetic fields, the liquid helium target is subdivided so that half the target is upstream from a $\pi$-coil and half downstream (see Fig.~\ref{target_idea}).  The 0.4\,mT vertical field of the $\pi$-coil reverses the direction of any horizontal rotation components from longitudinal fields by 180$^{\circ}$, thus undoing upstream rotations in the downstream half.  By filling only the upstream or the downstream half of the target with liquid, rotations from longitudinal fields are canceled while rotations due to the target are not.  Furthermore, the difference in measured rotations from the case of upstream filling versus downstream filling of the target halves results in the isolation of the rotation due to the target and the further suppression of rotations due to longitudinal fields.  While the $\pi$-coil is situated at the center of the four-quadrant target chamber, in practice it is not at the center of the low-field region between the transport coils.  Thus, rotations from the fields are not completely undone downstream of the $\pi$-coil.  We denote $x$ as the fraction of rotation that occurs upstream of the $\pi$-coil.  Another correction comes from the fact that the angle the $\pi$-coil rotates the neutrons about the vertical axis depends on the neutron velocity.  We denote $d$ as the velocity-dependent reduction in the analyzed polarization component.

To suppress sources of non-statistical noise, such as fluctuations in the neutron source intensity, fluctuations in magnetic fields, and electronic noise, the targets are further subdivided horizontally so that the apparatus functions as two side-by-side experiments, referred to as the east and west sides. The targets are filled so that when the upstream quadrant is filled on one side, the downstream side is filled on the other side.  Fig.~\ref{target_idea} shows conceptually how the diagonally-opposed target filling suppresses rotations from ambient fields while isolating the PV signal. The principle of measurement is discussed in detail in Ref.~\cite{Snow2015}

The four-target apparatus is machined out of a single piece of aluminum. Each quadrant has a length of 42\,cm equal to two mean-free paths of cold neutron scattering in liquid $^4$He  at 4\,K to optimize signal-to-noise ratio~\cite{Snow2015}. The fractional difference in length with other quadrants is $<10^{-4}$.  To handle the high count rates presented by the NG-6 beam, the $^3$He ionization chamber operates in current mode, integrating ionization charge initiated by neutron capture in $^3$He over a time interval which reduces electronic noise well below the statistical noise from neutron counts.  In addition, the ion chamber is segmented transversely into four quadrants to allow east-west, up-down signal separation.  It is also segmented longitudinally into four sections of increasingly longer lengths to capture roughly equal numbers of neutrons in four velocity classes based on the $1/v$ capture cross section of $^3$He in the energy range of the NG-6 beam~\cite{Penn2001}.  Because parity-violating rotations are velocity-independent while rotations due to magnetic fields depend directly on the time spent in the field, analysis of the rotations in each section leads to a measurement of residual magnetic fields in the target region and together with rotation measurements with artificially exaggerated fields provides a limit on systematic uncertainties related to magnetic fields.

\subsection{Data Products and Systematics Model}
The data acquisition is organized around various state changes of the apparatus starting with the flipping of the output coil field at 1\,Hz.  Every 10\,s the current in the $\pi$-coil is changed (on$+$, off, on$-$) and after 5 steps through the three $\pi$-coil states the target state is switched.  In this way, asymmetries and thus spin angles can be formed from count rate asymmetries for east and west sides of the polarimeter and both target states with the $\pi$-coil both off and on with reversed currents.
The spin angles may have contributions from misalignments in spin transport fields upstream of the target region and these may have dependencies on target configuration $\phi^{tr}(T)$. In addition there are spin precession angles from longitudinal magnetic fields found in the target region $\phi^m(T)$  and parity-violating spin rotation $\phi^{PV} = d\phi/dz \times L$, the rotary power per unit length times the length of liquid helium traversed by the neutrons. Spin angles for each state can be expressed in the following set of equations:

\begin{equation}
\phi_W(T_0) = d^W\phi^{tr}(T_0) + D^W(x) \phi^m_W(T_0) -  d^W\phi^{PV}
\end{equation}
\begin{equation}
\phi_E(T_0) = d^E\phi^{tr}(T_0)) +  D^E(x) \phi^m_E(T_0) +   \phi^{PV}
\end{equation}
\begin{equation}
\phi_W(T_1) = d^W\phi^{tr}(T_1) + D^W(x) \phi^m_W(T_1) +  \phi^{PV}
\end{equation}
\begin{equation}
\phi_E(T_1) = d^E\phi^{tr}(T_1) +  D^E(x) \phi^m_E(T_1) -  d^E \phi^{PV}.
\end{equation}
The terms $\phi_W$ and $\phi_E$ refer to the west and east sides of the polarimeter respectively and  $T_0$ and $T_1$ are the two target configurations. Neutrons pass through liquid helium before the $\pi$-coil in the west side polarimeter for target configuration $T_0$ and in the east side polarimeter for $T_1$. The factors $d^W$ and $d^E$ are the $\pi$-coil depolarization values for the two sides when the $\pi$-coil is on and the analyzing powers -$PA^0_W/PA^0$ and -$PA^0_E/PA^0$ when it is off.  $D^{W,E}(x)$ is an effective depolarization where $x$ is the fraction of rotation occurring upstream of the $\pi$ coil (see Eq.~\ref{eqn:eff}). The parity violating signal $\phi^{PV}$ is extracted from the following linear combination of these four angles,
\begin{equation}
\phi_{PNC} = \frac{\phi_W(T_0) - \phi_E(T_0) - \left(\phi_W(T_1) - \phi_E(T_1)\right)}{4},
\label{eqn:lincom}
\end{equation}
where $\phi_{PNC}$ is the net extracted angle measured by the apparatus.

 Equation~\ref{eqn:lincom} is a double subtraction of
west and east side polarimeter angles and target configurations. Substituting in the above set of spin angle expressions, each of the sequential target configurations can be written as shown here for $T_0$
\begin{eqnarray*}
\Delta\phi(T_0)&=&d^W\phi^{tr}(T_0) - d^E\phi^{tr}(T_0) + D^W(x) \phi^m_W(T_0)\\
&& -  D^E(x) \phi^m_E(T_0) - (1 + d^W) \phi^{PV}.
\nonumber
\end{eqnarray*}
Because the B field is nearly the same on both sides of the apparatus, $\phi^m_E(T_0) \approx \phi^m_W(T_0)$, we can define a west-east suppression factor $(1 - k^{0,\pi})/2$ which multiplies the average magnetic rotation angle $\phi^m$.
West-east differences for target configurations $T_0$ and $T_1$ are then given in equations~\ref{eqn:difT0} and \ref{eqn:difT1}. Thus, 

\begin{eqnarray}
\nonumber
\Delta\phi(T_0) &=& (d^W - d^E)\phi^{tr}(T_0) + (1 - k^{0,\pi})\phi^m(T_0)\\
&&  - (1 + d^W)\phi^{PV},
\label{eqn:difT0}
\end{eqnarray}

\noindent where $k^{0,\pi} \approx 1$ represents the degree to which analyzing powers on the two sides are equal. The superscripts $0$ and $\pi$ indicate whether the $\pi$-coil is off or on. Similarly, one can write the difference angle for the other target state

\begin{eqnarray}
\nonumber
\Delta\phi(T_1) &=& (d^W - d^E)\phi^{tr}(T_1) + (1 - k^{0,\pi})\phi^m(T_1)\\
&&  + (1 + d^E)\phi^{PV}.
\label{eqn:difT1}
\end{eqnarray}

\noindent Expressing Eq.~\ref{eqn:lincom} in terms of these target configuration differences shows how the measured spin rotation angle, $\phi_{PNC}$, depends on  transport field misalignments, magnetic fields, and parity-violating spin rotation.
\begin{eqnarray}
\nonumber
\phi_{PNC}&=&\frac{\Delta\phi(T_0) - \Delta\phi(T_1)}{4}\\
\nonumber
&=&(\frac{d^W - d^E}{2})(\frac{\phi^{tr}(T_0) - \phi^{tr}(T_1)}{2})\\ 
&+&(\frac{1 - j}{2})(\frac{1 - k^{0,\pi}}{2})\phi^m - \frac{1 + d}{2} \phi^{PV},
\label{eqn:thetapnc}
\end{eqnarray}

\noindent where  $d = (d^W + d^E)/2$ is the average depolarization, and the suppression factor written as  $(1 - j)/2$ represents the degree to which  $\phi^m$ is independent of target configuration.  The first two terms are major systematic contributions to the measurement. In this model transport field misalignment angles cancel when west and east side polarimeter responses are equal or if $\phi^{tr}$ is independent of target configuration. Spin precession from longitudinal magnetic fields $\phi^m$ is suppressed by a product of two factors. One would be null with equal angle responses in west and east polarimeters and the other if there were no dependence on target configuration. The third term is the parity violating signal extracted from the liquid targets and $\pi$-coil. In Section~\ref{sec:systematics}, we isolate magnetic field contributions to $\phi_{PNC}$ from transport misalignments and by replacing differences with sums (setting $j = -1$ or $k = -1$) we can turn off one or the other suppression factor, allowing  their values to be individually determined. 

\subsection{Crossed Polarizer-Analyzer Analyzing Power}
\label{sec:pap}

Obtaining spin angles from neutron count asymmetries  requires knowledge of the polarimeter's analyzing power given by the polarization product $PA$ of the polarizer-analyzer pair. Accurate determinations are especially important when computing angle differences that cancel common mode systematic effects. For a static system such as described here, $PA$ is stable to sufficient accuracy but was nonetheless measured periodically throughout the experiment by tilting the input coil by a few degrees and measuring the resulting asymmetry through the polarization analyzer.  In practice $PA$ differs slightly for different states of the $\pi$-coil and in particular depends on the reduction in the analyzed polarization component $d$ from the under-rotation and over-rotation of different velocity classes by the $\pi$-coil.  Rewriting Eq.~\ref{Asymmetry} to accommodate different measured neutron count asymmetries for the same rotation of the input coil in the case of $\pi$-coil on versus $\pi$-coil off gives

\begin{equation}
{{\cal A}}^0 = PA^0 \sin(\phi)\\
\label{eqn:pa0}
\end{equation}

\begin{equation}
{\cal A}^\pi = d \times PA^0 \sin(\phi),
\label{eqn:papi}
\end{equation}

\noindent where the $\pi$-coil on and $\pi$-coil off cases are indicated by superscripts $\pi$ and $0$, respectively, and $d=PA^{\pi}/PA^0$ gives the $\pi$-coil on response independent of polarizer-analyzer and spin transport efficiencies.

Neither $PA^{\pi}$ or $PA^0$ as defined above are sufficient to describe the analyzing power when internal magnetic fields are present. This is because the energized $\pi$-coil changes the sign of any horizontal rotation component acquired upstream of the $\pi$-coil and reduces the polarization projection by the factor $d$; rotations occurring downstream of the $\pi$-coil are added without this depolarization.  Thus, one arrives at a modified equation for the asymmetry

\begin{eqnarray}
\nonumber
{\cal A}^\pi(x)&=& PA^0 D(x) \sin(\phi)\\
&=& PA^x \sin(\phi),
\label{eqn:fldresp}
\end{eqnarray}
\begin{equation}
D(x) = \left(1 - (1 + d)x\right),
\label{eqn:eff}
\end{equation}

\noindent where $x$ is the fraction of the total rotation angle that occurred prior to the $\pi$-coil, and $d$ is the depolarization due to $\pi$-coil over-rotation or under-rotation.  The value of $PA^x$ is an effective polarization product for $\pi$-coil on data and is equivalent to the $\pi$-coil off value when $x = 0$. The small angle approximation for $\sin(\phi)$ is used in subsequent expressions with no loss in generality.

Reactor intensity fluctuations are a source of noise in asymmetry measurements common to both east and west side polarimeters. This noise is suppressed by an order of magnitude in taking polarimeter asymmetry differences \cite{Snow2015}.Ê   On the other hand, unequal west and east side values of $PA$, identified by $PA^{0,\pi}_W$ and $PA^{0,\pi}_E$,  degrade common mode angle performance when taking angle differences.  As an alternative to first converting individual asymmetries to angles we define multiplicative correction factors  to scale  west and east asymmetries to compensate their unequal analyzing powers. 

\begin{equation}
\omega^{0,\pi} = \sqrt{PA^{0,\pi}_W/PA^{0,\pi}_E},
\label{eqn:omega}
\end{equation}

\noindent Equations~\ref{eqn:pa0}, \ref{eqn:papi}, and \ref{eqn:omeg} show how applying these corrections to asymmetry differences recovers the original unbiased angle differences multiplied by the geometric mean of $PA$ products. If large enough however, these corrections can degrade the above mentioned reactor fluctuation noise performance. For our $PA$ product differences geometric and arithmetic averages are not significantly different:

\begin{equation}
\frac{1}{\omega}{\cal A_W} -  \omega {\cal A}_E = \sqrt{PA_W PA_E }\left(\phi_W - \phi_E \right).
\label{eqn:omeg}
\end{equation}

\begin{figure}
\includegraphics[height=8.5cm, clip]{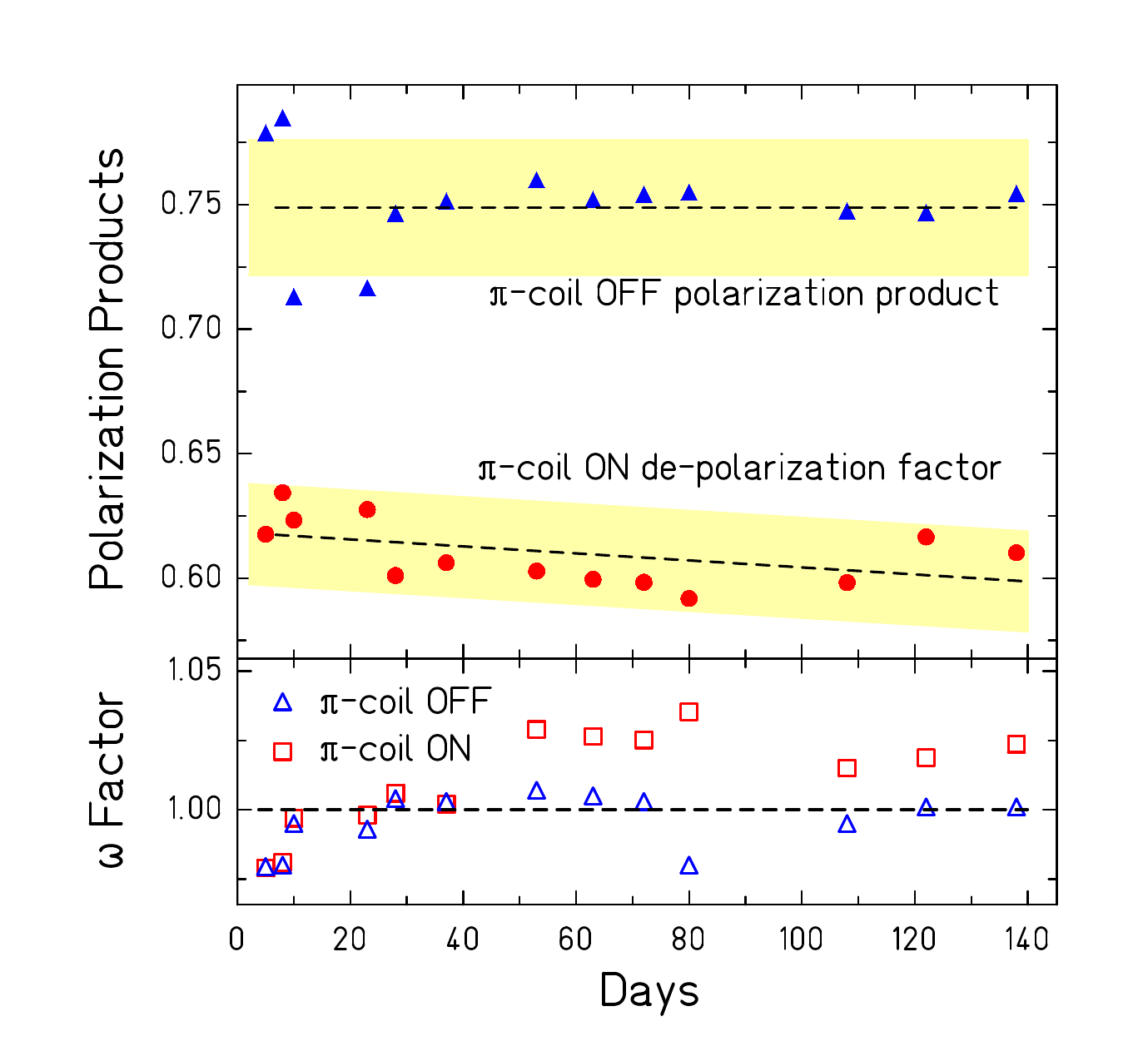}
\caption{The upper plot shows the average of the west and east polarimeter polarization products $PA$  with the $\pi$-coil off and corresponding de-polarization factors $d$ when the $\pi$-coil was on~\cite{Snow2015}. The error bars on individual points represent one-sigma, statistical uncertainties; shaded bands show average errors per point based on their means. The lower plot gives $\omega^0$ (triangles) and $\omega^{\pi}$ (squares), the fraction by which each side differs from the average for polarization products and de-polarizations respectively.}
\label{fig:ppt}
\end{figure}

\noindent The upper plot in Fig.~\ref{fig:ppt} shows the 13 independently measured values for $PA^0$ and $\pi$-coil de-polarizations $d$ averaged over both target configurations over the course of the experiment. The fractional errors on these measurements were 1.6\%. The lower plot gives $\omega^{0,\pi}$ corrections to west and east side asymmetries. The data analysis used asymmetries scaled by the closest preceding $PA$ products to determine spin angles.

\section{Experimental determination of magnetic field-correlated systematic effects}
\label{sec:systematics}

The extraction of systematic uncertainty from the experimental data proceeds through the following steps.  First we isolate contributions to the total spin rotation angle from the different sources given in Eq.~\ref{eqn:thetapnc}. We use the ion chamber calibration of the polarimeter's sensitivity to longitudinal magnetic fields along with the measurements of polarization products for the two sub-beams to experimentally determine the degree of common-mode suppression of the spin rotation from internal magnetic fields. One then multiplies this suppression factor by the value of the average field to extract the systematic component in the measurement.

We use data taken with the $\pi$-coil off to measure longitudinal magnetic fields in the storage region. Because the ion chamber is wavelength sensitive (downstream longitudinal sections having a harder spectrum) and the rotation angle due to traversing a longitudinal magnetic field $B_L$ for a distance $L$ depends linearly on wavelength
\begin{equation}
\phi_m=\frac{m_n \gamma_n B_L L \lambda}{\hbar},
\end{equation}
one expects the rotation angle to decrease for signals from sections further into the ion chamber.  By applying an artificially enhanced magnetic field alternating between  plus and minus 0.5\,$\mu$T we calibrated how the average rotation from magnetic fields depends on the fractional change in rotation angle as a function of longitudinal plane in the ion chamber. From the slope of this dependence, rotations due only to magnetic fields can be obtained from $ (\phi_M)^{-1}d\phi_{PNC} / dP = (-6.96\pm 0.09)\times 10^{-2}$  \cite{Snow2015}. With the $\pi$-coil energized this changes to $ (\phi_M)^{-1}d\phi_{PNC} / dP = (-32.5\pm 0.26)\times 10^{-2}$. From the ratio of these calibration constants we see the effect of the energized $\pi$-coil is to reduce the size of magnetic field caused rotations by a factor of 0.21. As described in Section~\ref{sec:exp}, this results from an {\it in situ} cancellation and gives rise to the effective $PA$ product in Eq.~\ref{eqn:fldresp}.  Solving for $x$ with this reduction factor and the average depolarization $d$ of 0.6 gives $x = 0.49$, essentially the physical location of the $\pi$-coil at target center ($x = 0.5$). Angles resulting from transport field misalignments or parity violation are independent of ion chamber plane and do not contribute in these expressions. Isolating rotations from  magnetic fields in this manner allows us to re-write Eq.~\ref{eqn:thetapnc} in terms of the above $\phi_M$ which then simplifies to
\begin{equation}
\phi_M = (\frac{1 - j}{2})(\frac{1 - k^{0,\pi}}{2})\phi^m. 
\label{eqn:thetaM}
\end{equation}
Replacing differences with sums in the equation for extracting $\phi_{PNC}$ (Eq.~\ref{eqn:lincom}) corresponds to setting the $j$ or $k$ parameter to -1 in the above expression, which turns off the corresponding suppression factor. The $\phi_M$ angle ratio of suppression-on to suppression-off gives the magnitude of the corresponding suppression factor. Averages of the side-by-side polarimeter angles lack the reactor intensity fluctuation suppression found in their differences.  In determining the k suppression factor, the better signal-to-noise magnetic field data from the ion chamber calibration was used for $\phi^m$ which has by default no target suppression. For $\pi$-coil off data the suppression factor was  $(356)^{-1}$.  For $\pi$-coil on data, these values were $(270)^{-1}$ correcting asymmetries for $\omega^0$ and $(134)^{-1}$ correcting for $\omega^{\pi}$. 
 
Target suppression factors are determined by turning off east-west suppression where $\phi^m$ comes from the full run data set with the $\pi$-coil off. $\phi_M$ with target suppression on is  $-(2.59 \pm 0.46) \times 10^{-5}$\,rad with the $\pi$-coil off and $-(1.30 \pm 0.11) \times 10^{-5}$\,rad with the $\pi$-coil on. With no target suppression, the $\pi$-coil off angle $\phi_M$ is $(1.0 \pm 0.14) \times 10^{-2}$\,rad, which is the average longitudinal magnetic spin precession angle over the run.  Dividing by the response to a known longitudinal magnetic field, measured to be $(2.38 \times 10^{-1})$\,rad/$\mu$T, gives $(42 \pm 6)$\,nT.  The ratios of suppression-on angles to the $\pi$-coil off angle without suppression gives target suppression factors  $-(386 \pm 64)^{-1}$ with the $\pi$-coil off and $-(767 \pm 54)^{-1}$ with the $\pi$-coil on.

Using these values, the combined suppression of longitudinal magnetic field uncertainty from both subtractions  is given by
\begin{eqnarray*}
 \mbox{ $\pi$-coil off} &=&(-7.0 \pm 0.2) \times 10^{-6}\\
 \mbox{$\pi$-coil on( $\omega^0$)} &=&(-4.8 \pm 0.3) \times 10^{-6}\\
 \mbox{$\pi$-coil on( $\omega^{\pi}$)} &=&(9.7 \pm 0.6) \times 10^{-6}.
\nonumber
\label{eqn:totsupp}
\end{eqnarray*}

\noindent The run average magnetic precession angle $(1.0 \times 10^{-2})$\,rad times the suppression factor gives the magnetic systematic remaining in $\phi_{PNC}$.  For $\pi$-coil off data this is ($-7.0 \pm 0.2) \times 10^{-8}$. For $\pi$-coil on data these values are $(-4.8 \pm 0.3) \times 10^{-8}$\,rad for $\omega^0$ and $(9.7 \pm 0.6) \times 10^{-8}$\,rad for $\omega^{\pi}$.

\begin{figure}[ht]
\includegraphics[height=6.2cm, clip]{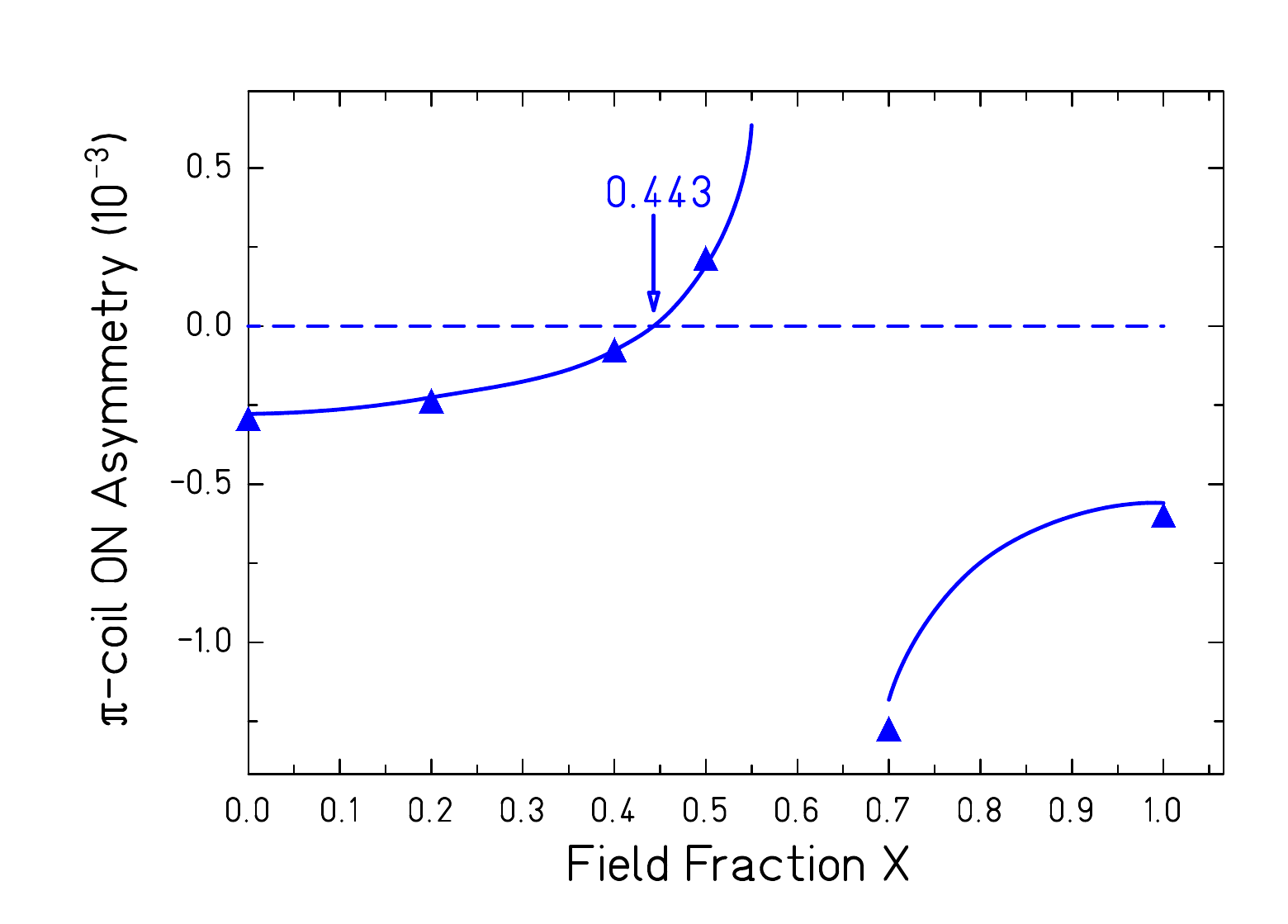}
\caption{West and east side asymmetry differences are shown corrected for equal responses to uniform longitudinal fields. The fraction of the total rotation that occurs before the $\pi$-coil is given by $x$. The curve represents a spline fit to the data points and crosses zero when the two sides have equal responses to the longitudinal field.}
\label{fig:tcresponse}
\end{figure}

The same calibration data set can be used to measure how well the polarization product compensation works in common mode  cancellation of magnetic systematic rotations in $\phi_{PNC}$.
West and east side $\pi$-coil on asymmetries are corrected for their unequal $PA$ products using Eq.~\ref{eqn:omega} and the effective polarization product $PA^x$ from Eq.~\ref{eqn:fldresp}.  The difference of west and east asymmetries is shown in Fig.~\ref{fig:tcresponse} for values of $x$ that were chosen to span the range $0 \rightarrow 1$.  $PA^x$ becomes zero near $x \approx 0.6$ and gives rise to the pole in the plot.  Both sides measure the same magnetic angle for the value of $x$ that best matches the polarimeter's actual response.  The solid lines are spline fits to the points and show the asymmetry difference curve going through zero at $x = 0.443$. 

\begin{table}
\caption{ Values of $\phi^M_{PNC}$ have west and east asymmetries corrected for equal $PA$ products. For $x = 0$ and $x = 1$, corrections are for equal $\pi$-coil  off and $\pi$-coil  on $PA$ products, respectively. At $x = .443$ the correction is for equal responses to uniform longitudinal fields.}
\begin{tabular}{lcc}
\hline\hline
Correction		& Fraction 	& $\phi^M_{PNC}$ 				\\
\hline
$PA^x$		&  $x = 0.443$	& $(0.047 \pm .016)\times 10^{-4}$ 	\\
$PA^x$		&  $x = 0$		& $(-4.44 \pm .016)\times 10^{-4}$ 	\\
None		& 			&  $(-5.80 \pm .016)\times 10^{-4}$ 	\\
$PA^x$		& $x = 1$		&  $(-8.95 \pm .016)\times 10^{-4}$ 	\\
\hline\hline
\label{tab:magtheta}
\end{tabular}
\end{table}

In this data set the target configuration is not changed, and target states $T_0$ and $T_1$ are instead associated with the sign of the externally applied uniform magnetic field. 
 In Eq.~\ref{eqn:fldresp}, $x = 0$ and $1$ correspond to corrections for equal $\pi$-coil off  and on responses respectively. Table~\ref{tab:magtheta} gives the corresponding $\phi^M_{PNC}$ values for $x = 0, 0.443, 1$, and the case where no response function compensation was applied. The ion-chamber calibrations using purely magnetic rotations placed $x$ at the center of the target. The offset from target center measured here for $\phi^M_{PNC}$ results from transport misalignment systematics not canceled with this polarimeter compensation value. The table also shows that when suppressing magnetic rotations, compensation by $\omega^{\pi}$ is worse than $\omega^0$  by a factor of 2. This same relationship was seen in the measured suppression factors previously discussed.

\section{Data Analysis}
\label{sec:anal}

In Section~\ref{sec:systematics} using neutrons as a co-magnetometer, we obtained the mean longitudinal magnetic field in the target region and corresponding systematic spin rotation contribution to $\phi_{PNC}$. Another systematic contribution was found from target dependent changes in the misalignment of the neutron spin angle at the input to the target region $\phi^{tr}$. The analysis methods that follow deal with these systematics at different time scales. The first computes average spin angles and their uncertainties for each individual run  sequence and then forms a weighted mean of their values. This method samples magnetic field variations at roughly 8 hour intervals. The second method computes $\phi_{PNC}$ for each pair of target configurations and forms a global mean weighting all points equally, sampling magnetic field variations at intervals of order minutes. The results for the parity-odd spin rotation angle and the systematic uncertainty correlated with internal magnetic fields are presented.  We begin with a description of the various cuts that were applied to the data.

\subsection{Cuts Applied to the Data}

\begin{figure}
\includegraphics[height=7.7cm, clip]{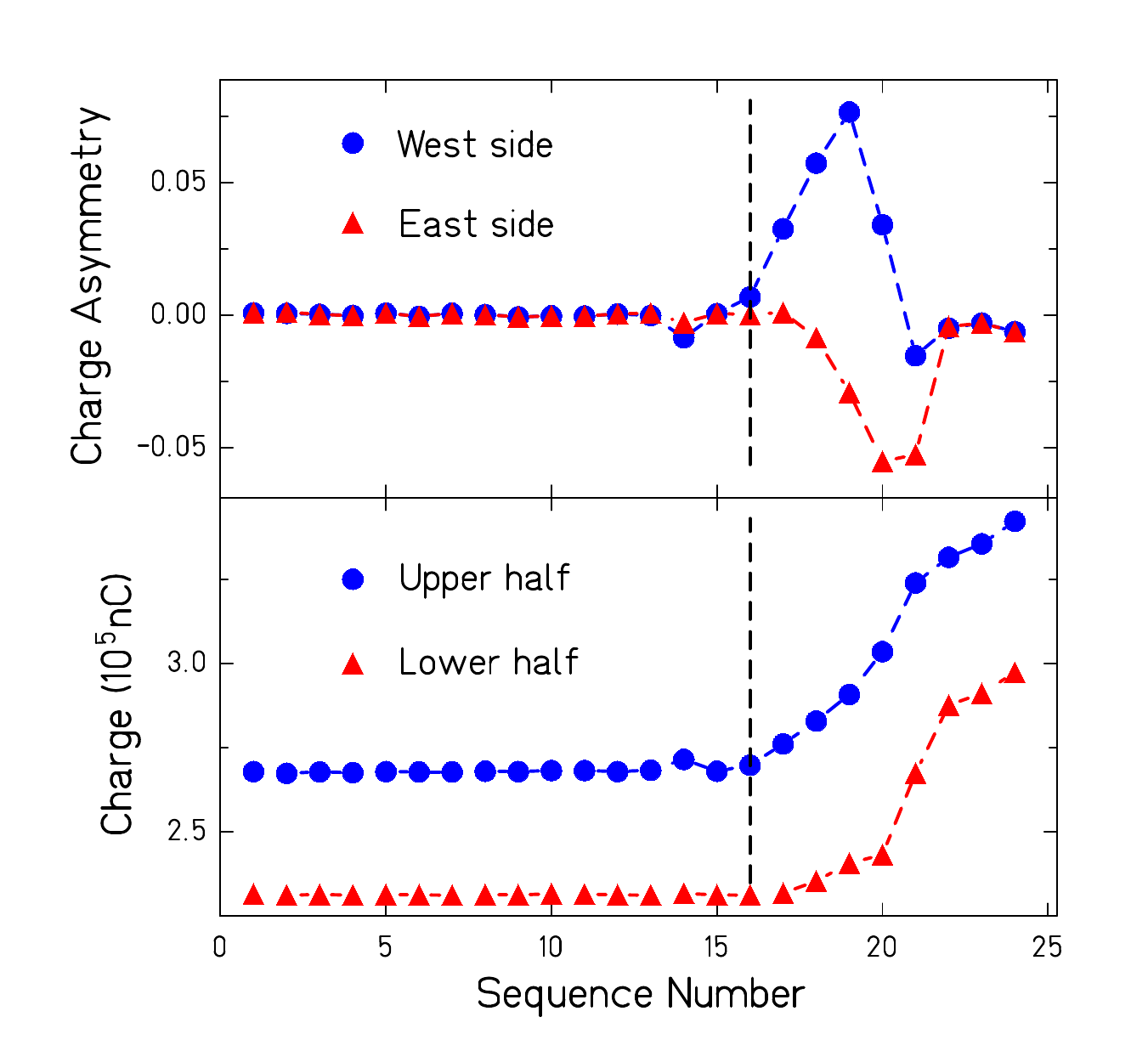}
\caption{The upper plot shows the fractional difference in charge collected in the ion chamber between configurations with liquid in the upstream target and the downstream target, for both west and east sides of the beam. This data gives time derivatives of liquid levels for each sequence. The lower plot gives the total charges collected from the upper and lower halves of the target. The vertical dashed line shows where the pump could no longer fill the targets as the target chamber runs out of liquid helium.} 
\label{fig:dutyF}
\end{figure}

Run durations were each about 8 hours, set by the hold time for the liquid helium in the target vessel. Run data was visually inspected for incomplete filling of the target chambers as it is essential that the full cross sectional area of both halves of the neutron beam see only liquid helium. Liquid helium attenuates the neutron flux much more than gaseous helium so the total charge collected for each sequence is an excellent proxy for the liquid level. Figure~\ref{fig:dutyF} shows data where the pump could no longer completely fill the targets near the end of the run as the liquid level in the vessel falls too low. The upper plot shows target charge asymmetries $(N^{Tot}_{T_0} - N^{Tot}_{T_1})/(N^{Tot}_{T_0} + N^{Tot}_{T_1})$ for each side of the experiment plotted against sequence number. The lower plot shows corresponding integrated charges for upper and lower halves of the detector. In this example continuously dropping liquid levels are clearly visible after sequence 16. Run sequence values greater than 16 (dashed vertical line) were therefore excluded. A total of 3102 out of 4107 sequences were used in our analysis after this cut was applied amounting to a 25\% deadtime.

\subsection{Run-by-Run Analysis}\label{sec:anal_avg}


Data from the three $\pi$-coil states are analyzed independently. The two $\pi$-coil on polarities are then averaged to cancel possible systematic effects linear in the $\pi$-coil current, which might come from the small but nonzero external magnetic fields of the $\pi$-coil. Spin angles are computed for west and east side polarimeters for each target configuration in the sequence. The spin rotation angle $\phi_{PNC}$ if extracted directly from Eq.~\ref{eqn:lincom}, can be biased by time-dependent variations in the longitudinal magnetic field on timescales long comparable to the liquid motion frequency.  We instead use the method given in Ref.~\cite{Swanson2008} to remove any slow (zero-point) drifts from this data. A short description of the procedure is outlined in the steps below.

For each run sequence, a time series $u_i$ is constructed of west and east spin angle differences for alternating target configurations. This series has the form $\{A,B,A,B,\ldots\}_{\in N}$, where $A$ and $B$ correspond to target configurations $T_0$ and $T_1$,Ê respectively, and $N$ is twice the number of sequences in a run. The time series is then convoluted with a filter algorithm that removes constant and linear and quadratic time-dependent variations according to the transformation

\begin{equation}
y_i  = \frac{(-1)^{i}}{8}\left( u_i -3u_{i + 1} + 3u_{i + 2} - u_{i + 3}\right).
\end{equation}

With this combination of the data, differences in successive target configurations are preserved and the double subtraction in Eq.~\ref{eqn:lincom} is recovered. The factor $(-1)^i$ is included to demodulate target configuration-dependent components in the new series in order to determine their mean value. The $u_i$ terms are the original time series data and the $y_i$ terms form a new magnetic field drift-free data set. The convolution is expressed in matrix notation as a linear transformation between these two sequences

\begin{equation}
Y  = G U,
\end{equation}

\noindent where $Y$ is a column vector of length $N - 3$ containing the filtered data, $U$ is a column vector containing the original data of length $N$, and $G$ is an $N \times (N-3)$ matrix containing the filter coefficients. The mean and standard deviation of the filtered data set are computed using the covariance matrix from the convolution.  The original data set was assumed to be uncorrelated with diagonal covariance matrix $Cov(U,U)$, an $N \times N$ matrix with elements $(\sigma^2_{u_i})_{i,i}$. The corresponding covariance matrix for the filtered set is given by

\begin{equation}
C = Cov(Y,Y) = G Cov(U,U)G^T.
\end{equation}

The parity-odd signal $\phi_{PNC}$ is the mean of the filtered data $Y$, which is obtained by weighting the $y_i$ elements by the inverse covariance matrix and computing a weighted average

\begin{equation}
\mu = \frac{X^T C^{-1}}{X^T C^{-1}X} Y,
\end{equation}

\noindent where the design matrix $X$ is a column vector of length $N-3$ with all elements $= 1$. The variance in $\mu$ is given by

\begin{equation}
\sigma^2_{\mu} = \frac{1}{X^T C^{-1}X},
\end{equation}

\noindent the inverse of the sum of all elements of the covariance matrix. Individual run means are then combined using inverse square uncertainty weighting to obtain the final result. 

Using this method independent analyses were carried out with west and east side $PA$ product corrections as given by Eq.~\ref{eqn:omega}. For $\pi$-coil on data, rotations occurring primarily upstream of the coil required $\omega^{\pi}$ polarimeter corrections while those downstream required $\omega^{0}$. For comparison analyses were also performed for $\pi$-coil on data without corrections and for $\pi$-coil off data.   The results from these analyses are shown in Table~\ref{tab:theta}. 
\begin{table}
\caption{ $\phi_{PNC}$ is given for different polarization product corrections. Column 4 gives the $\chi^2/dof$ of the weighted means of the run sets.}
\begin{tabular}{lccc}
\hline\hline
$\pi$-coil	&	Correction		& $\phi_{PNC} (\rm{rad})$ 			& $\chi^2/dof$\\
\hline
On 		&    $\omega^{\pi}$	&  $(0.7 \pm 2.8)\times 10^{-7}$	& 283/234\\
On 		&    $\omega^0$	&  $(-2.2 \pm 2.6)\times 10^{-7}$	& 320/234\\
On 		&    None			&  $(-3.2 \pm 2.6)\times 10^{-7}$	& 337/234\\
Off 		&   $\omega^0$		&  $(-0.6 \pm 3.8)\times 10^{-7}$	& 283/234\\
Off 		&   None			&  $(-2.6 \pm 3.7)\times 10^{-7}$	& 294/234\\
\hline\hline
\label{tab:theta}
\end{tabular}
\end{table}

\subsubsection{Discussion of Run-By-Run Results}

The weighted  mean of all run sets is shown in column 3. Except for the first row, uncertainties are a few percent above the neutron shot noise. The size of the $\pi$-coil $\omega^{\pi}$ corrections in row 1 result in slightly worse noise performance from reactor intensity fluctuations. 

The two types of direct systematic contributions to the measured angle $\phi_{PNC}$ considered were  rotations from target configuration dependent neutron transport fields at the input to the target region and rotations from longitudinal magnetic fields in the target region. From Eq.~\ref{eqn:thetapnc}, contributions from the first type are best canceled when west and east side polarimeters have equal analyzing powers. As previously discussed, the $PA$ products for west and east polarimeters can be compensated by applying correction factors to the count rate asymmetries. When the $\pi$-coil is off there is no contribution from parity-odd spin rotation, so $\phi_{PNC}$ should be zero in the absence of any systematic rotations. The $\pi$-coil off entries in Table~\ref{tab:theta} show how compensating $PA^0$ products brings this angle closer to zero. Reducing contributions of this systematic type with the $\pi$-coil on requires $\omega^{\pi}$ corrections to the asymmetry data as these rotations occur upstream of the $\pi$-coil. This analysis result is given in row 1 of Table~\ref{tab:theta}. A comparison with the uncorrected result in row 3 gives the size of this systematic contribution  before cancellation as $3.91\times 10^{-7}$\,rad.

In Section~\ref{sec:systematics} we saw that systematic contributions of the second type could be suppressed but not completely canceled using either $\omega^{\pi}$ or $\omega^0$.  The size of the uncanceled contribution determined in Section~\ref{sec:systematics} by multiplying the average magnetic precession angle by the measured suppression factors was found to be $(9.7 \pm 0.6) \times 10^{-8}$\,rad. To verify the arithmetic sign of the rotation, estimators of correlations between individual run $\phi_{PNC}$ values and longitudinal field measurements were calculated

Variations in individual run mean values are greater than mean errors as indicated by chi square values in Table~\ref{tab:theta}. The longitudinal magnetic field in the target region fluctuated around a value of $-50$\,nT for the first two reactor cycles, then changed sign and varied around  $+50$\,nT in the third cycle. Mean values for each individual run set include unsuppressed spin precessions from longitudinal fields at the time of the run. Magnetic spin precession is linear in the magnetic field so the uncanceled magnetic spin angle in the weighted mean of all run sets correctly scales with the average magnetic field over the entire running period.  The variance in the mean is reflected in the observed $\chi^2/dof$ values in Table~\ref{tab:theta}. Typical magnetic field fluctuation spectra follow 1/f distributions and would lead to an increased variance on the time scales of a full run sequence.

\begin{figure}
\includegraphics[height=5.2cm]{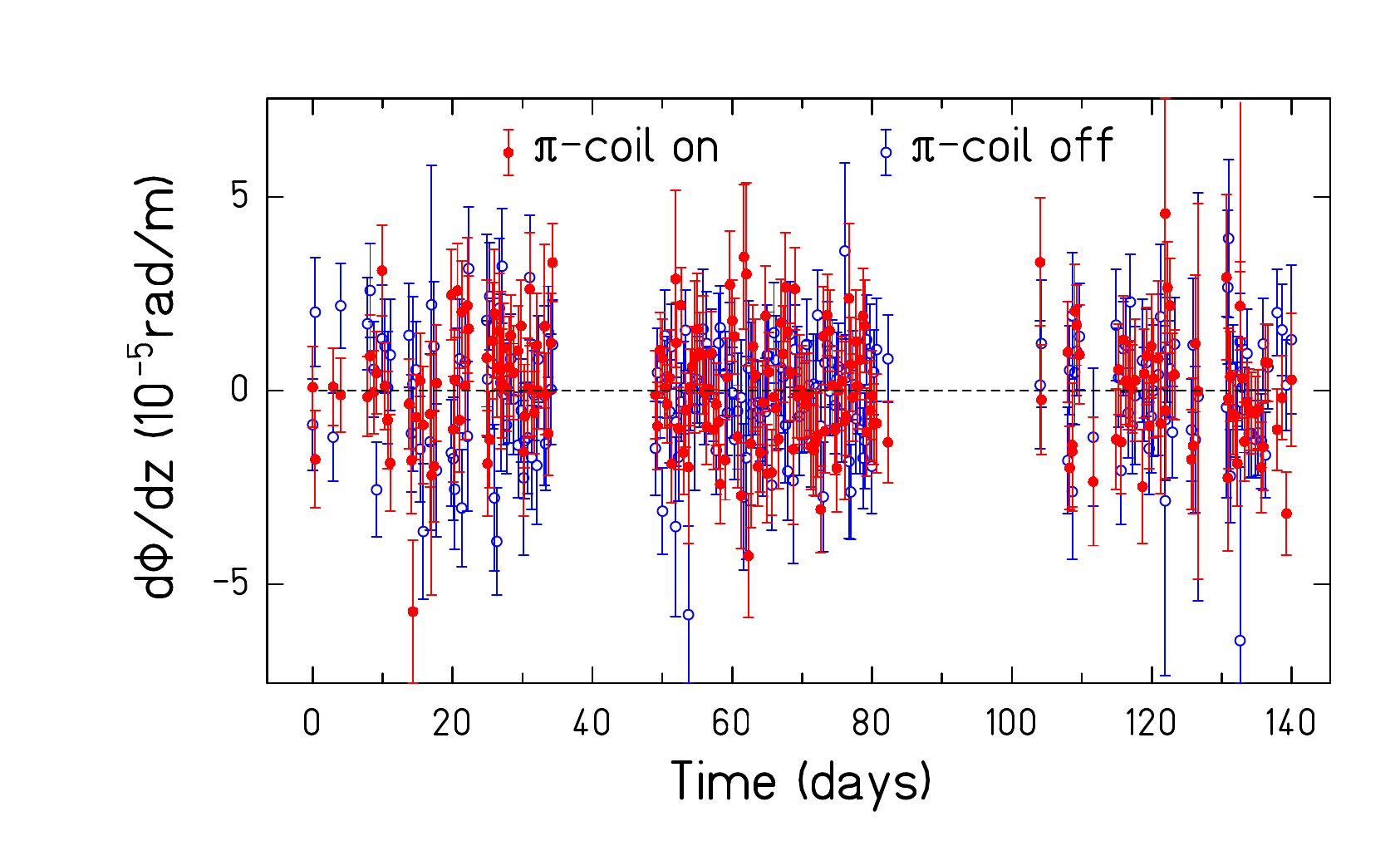}
\caption{$d\phi/dz$ determinations are shown for each run in the 3 reactor cycles. $\pi$-coil on values have filled circles (red on line) and $\pi$-coil off values have open circles (blue on line). The error bars represent one-sigma statistical uncertainty.}
\label{fig:pvangles}
\end{figure}

Our measured result from Table~\ref{tab:theta} is $\phi_{PNC} = (0.72\pm 2.8(stat.)\, ^ {+0.97} _{-0.07} (sys.)) \times 10^{-7}$\,rad, where the uncanceled magnetic rotation is accounted for in the systematic uncertainty. To express  our measured $\phi_{PNC}$  value as $d\phi/dz$ in units of radians/meter, it must be corrected for $\pi$-coil de-polarization (see Eq.~\ref{eqn:thetapnc} where for $d = 0.6$ the correction is 0.80) and scaled to the length (42\,cm) of the liquid helium target. This yields the result
\begin{equation}
\frac{d\phi}{dz} = (2.1 \pm 8.3(stat.)\, ^ {+2.9} _{-0.2} (sys.)) \times 10^{-7} \mbox{rad/m}.
\end{equation}
\noindent The systematic uncertainty is asymmetric because the effect from the precession from the residual longitudinal magnetic field is only positive. The individual run $d\phi/dz$ values for the three reactor cycles are shown in Fig.~\ref{fig:pvangles}.

\subsection{Point-by-Point Analysis}
\label{sec:ana_pbpl}

For the results discussed above the data were analyzed on a run-by-run basis: one produces a rotation angle after every complete sequence of the data acquisition. A sequence consists of a series of time-ordered measurements of east and west spin angle differences for both target configurations $T_0$ and $T_1$, as presented in Section~\ref{sec:anal_avg} and discussed in detail in Ref.~\cite{Snow2015}. The analysis combined all the changes of the state of the output coil $N^+$ and $N^-$ for given $\pi$-coil and target states. It is also possible to extract rotation angles using the minimum amount of data by requiring only the two changes of state of the output coil for both target states in a given $\pi$-coil state. This produces a number of independent measurements of the rotation angle on a point-by-point basis. The total number of measurements is equal to the product of (total number of sequences)*(number of rotation flips of the output coil)*(number of $\pi$-coil sequences), which is $(3136*10*5)=$ 156,850 for this data set.

\begin{figure}[t]
\includegraphics[width=3.4in]{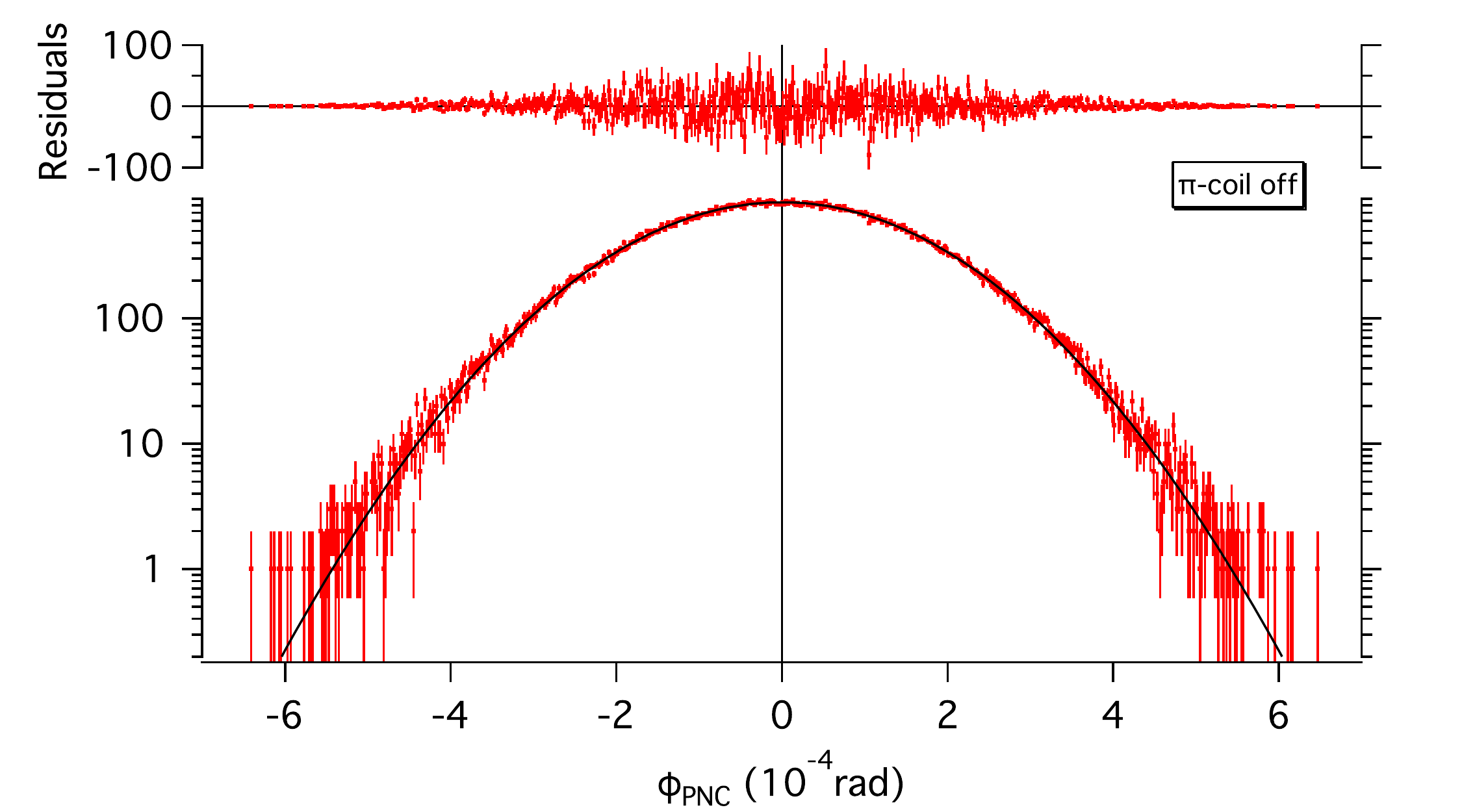}
\includegraphics[width=3.4in]{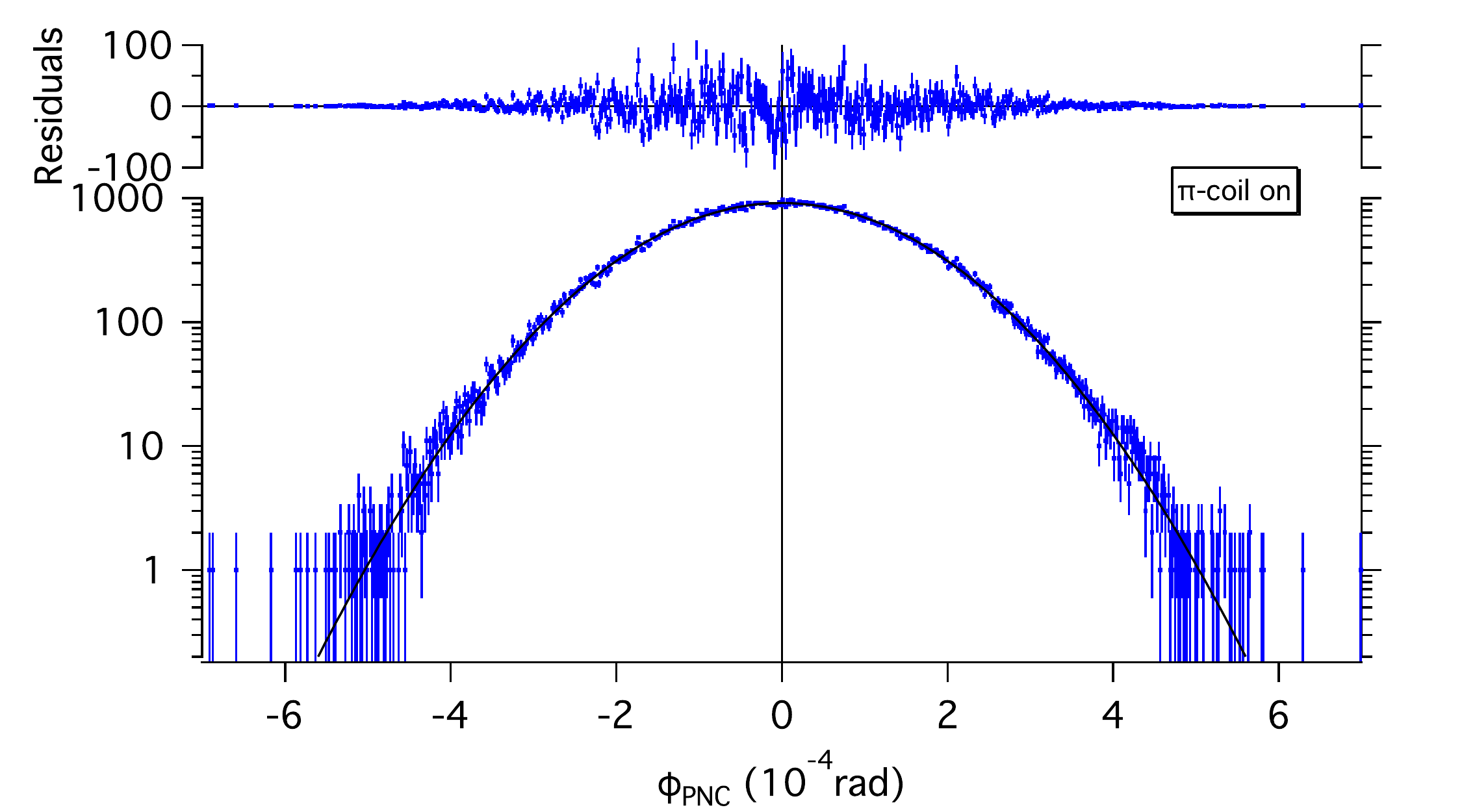}
\caption{\label{fig:pbp} Distribution of measured spin rotation angles in liquid $^4$He with $\pi$-coil off (upper plot) and $\pi$-coil on (lower plot).  The error bars represent one-sigma statistical uncertainty. The solid lines are fits to a Gaussian function, and the residuals are shown above the data. }
\end{figure}

The rotation angle data can be obtained from this point-by-point analysis method by determining the peak position from a histogram of the points.  The value of such an analysis is that it serves as a strong check on the sequence analysis. It permits a check not only on the central value of the rotation angle but allows a high fidelity search for systematic effects that may produce asymmetries in the histogram. Figure~\ref{fig:pbp} shows the data for the $\pi$-coil on and off analyzed in the point-by-point method. The histograms of the distribution of rotation angles were fit to a Gaussian function with no background term. For the $\pi$-coil off data, the fit gives a central value of $\phi_{PNC} = (-0.4 \pm 3.8)\times 10^{-7}$\,rad with a $\chi^2{\rm} /dof =512/566$; this value can be compared with the results from the run analysis given in Table~\ref{tab:theta}. For $\pi$-coil on data, the central value after correcting for the target length  is $d\phi/dz = (+1.6 \pm 8.2)\times 10^{-7}$\,rad/m with a $\chi^2{\rm}/dof = 566/536$. 

These results contain the uncanceled magnetic precession systematic measured at shorter time intervals than in the run-by-run analysis. The smaller values of $\chi^2{\rm}/dof$ result from sampling the fluctuation power spectrum at higher frequencies. These two analysis methods use essentially the same data set but are analyzed in ways that are fundamentally different. One coarsely bins filtered data run-by-run into runs and takes their weighted average. The other histograms all sequence points directly and fits them to a Gaussian. These approaches have different sensitivities to systematic errors from the binning and weighting. The difference of $0.5 \times 10^{-7}$ in mean values is well within our stated statistical uncertainty. Furthermore, to demonstrate that our analysis methods were able to extract parity violating spin rotations, we injected a simulated angle of $1 \times 10^{-5}$\,rad along with Gaussian noise into a subset of the data where only cold helium gas was present. Both the run-by-run and point-by-point analysis methods correctly extracted this angle from the data. For the average number of sequences in the run-by-run analysis~\cite{Swanson2008}, one estimates a 1.5\% increase in uncertainty over Gaussian statistics. This is in good agreement with statistical uncertainties in the two determinations.

\section{Experimental Uncertainty}
\label{sec:uncert}

Uncertainties in the angle measurement can be organized into three classes: (a) multiplicative effects, which affect the absolute size of a true parity-violating rotation, (b) non-statistical random uncertainty from fluctuations in measurement parameters, and (c) rotations arising from sources other than parity violation that produce systematic errors. Class (b) uncertainties are mitigated by subtracting rotation angles from the east and west sides to remove common-mode noise. This procedure results in a factor of about 10 reduction in non-statistical random uncertainty leading to a total random uncertainty that is 1.8\% larger than $\sqrt{N}$. The resulting statistical uncertainty is close to that expected from neutron counting statistics~\cite{Snow2015}.  

The potential systematic errors of class (c) are varied and have been investigated through calculation, simulation, auxiliary measurements~\cite{Snow2015} and this analysis.  Table~\ref{tbl:sys} enumerates estimates of the size of potential systematic uncertainties for this experiment from calculation and simulation and shows that all are at the $10^{-7}$ level of the experimental goal of this measurement. These sources of systematic uncertainty were discussed in detail in~\cite{Snow2015} and are not repeated here.

\begin{table}
\caption{A list of sources for potential systematic effects and estimates for the uncertainties~\cite{Snow2015}. The values for the uncertainties originate from either a calculation or are the result of a direct measurement that places an upper bound on the effect.}
\begin{center}
\label{tbl:sys}
\begin{tabular}{lcc}
\hline\hline
 Source                    				&   Uncertainty (rad/m) 	& Method \\
 \hline
  liquid $^{4}$He diamagnetism		& $2\times 10^{-9}$		&calc. \\
  liquid $^{4}$He optical potential	& $3\times10^{-9}$		&calc. \\
  neutron E spectrum shift			& $8\times10^{-9}$		&calc. \\
  neutron refraction/reflection		& $3\times10^{-10}$		&calc. \\
  nonforward scattering			& $2\times 10^{-8}$		&calc. \\
 uncanceled B field				& $2.9\times10^{-7}$		&meas. \\
\hline\hline
\end{tabular}
\end{center}
\end{table}

\section{Conclusion}
\label{sec:concl}

Our result for the neutron spin rotation angle per unit length in $^{4}$He of  $d\phi/dz =(+2.1 \pm 8.3 (stat.) \, ^ {+2.9} _{-0.2}  (sys.))\times10^{-7} $\,rad/m is consistent with zero and supersedes the result from Refs.~\cite{Snow2011,Snow2015}. Although this value is marginally different from that published in 2011, it represents a more sophisticated treatment of the systematic effects of the residual magnetic field and emphasizes the need to improve both the polarizer/analyzer uniformity and the magnetic shielding in any future measurement. We have modeled systematic contributions to the measured result in two classes: those dependent on neutron wavelengths and those which are independent. Wavelength analysis of the data allowed us to determine the precession angle due to longitudinal magnetic fields in the target region. Interpreted as $d\phi/dz$, this was found to be $2.9 \times 10^{-7}$ rad/m. Compensating the non-uniformity of $PA$ products canceled the other class of systematic angles by common mode suppression.

A second phase of the measurement is planned at a more intense neutron beam constructed at NIST~\cite{Cook09}. Improvements to the apparatus include better-optimized magnetic shielding and control of external field fluctuations,  a neutron polarizer and analyzer with improved phase space uniformity, and nonmagnetic supermirror input and output guides all of which will further reduce the systematic uncertainty, which can then be experimentally verified.  The apparatus will incorporate an improved liquid helium pump and a helium liquefier to reduce deadtime. With these improvements, we expect to reduce the statistical uncertainty on $d\phi/dz$ to better than  $2\times 10^{-7}$\,rad/m with smaller systematic uncertainties. If the prediction of $d\phi/dz = (9 \pm 3)  \times 10^{-7}$\,rad/m from the $1/N_{c}$ analysis as applied to NN weak amplitudes is correct, then a nonzero parity-odd neutron spin rotation in $\vec{n}+^{4}$He would be clearly resolved and would constitute to our knowledge the first successful prediction of a NN weak amplitude from the Standard Model. 

\section{Acknowledgments}

This work was supported in part by the National Science Foundation grants NSF PHY-0457219 and NSF PHY-0758018 and Department of Energy grants DE-AI02-93ER40784, DE-FG02-95ER40901, DE-SC0010443, and DE-FG02-97ER41020 MOD 0053. We acknowledge the support of the National Institute of Standards and Technology, US Department of Commerce, in providing the neutron facilities used in this work. WMS acknowledges support from the Indiana University Center for Spacetime Symmetries.

\newpage
\section*{References}
\bibliographystyle{apsrev}
\bibliography{NSRPRC}
\end{document}